\documentclass[aps,prd,twocolumn,groupedaddress,showpacs,nofootinbib]{revtex4}
\usepackage{graphicx,amssymb}

\begin{document}

\title{Reconciling dark energy models with $f(R)$ theories}

\author{S. Capozziello}
\author{V.F. Cardone}
\thanks{Corresponding author, email: {\tt winny@na.infn.it}}
\author{A. Troisi}
\affiliation{Dipartimento di Fisica ``E.R. Caianiello'', Universit\`a di Salerno
and INFN, Sez. di Napoli, Gruppo Coll. di Salerno, via S. Allende, 84081 - Baronissi (Salerno), Italy}

\begin{abstract}

Higher order theories of gravity have recently  attracted a lot of interest as alternative candidates to explain the observed cosmic acceleration without the need of introducing any scalar field. A critical ingredient is the choice of the function $f(R)$ of the Ricci scalar curvature entering the gravity Lagrangian and determining the dynamics of the universe. We describe an efficient procedure to reconstruct $f(R)$ from the Hubble parameter $H$ depending on the redshift $z$. Using the metric formulation of $f(R)$ theories, we derive a third order linear differential equation for $f(R(z))$ which can be numerically solved after setting the boundary conditions on the basis of physical considerations. Since $H(z)$ can be reconstructed from the astrophysical data, the method we present makes it possible to determine, in principle,  what is the $f(R)$ theory which best reproduces the observed cosmological dynamics. Moreover, the method  allows to reconcile dark energy models with $f(R)$ theories finding out what is the expression of $f(R)$ which leads to the same $H(z)$ of the given quintessence model. As interesting examples, we consider "quiessence" (dark energy with constant equation of state) and the Chaplygin gas.

\end{abstract}


\maketitle

\section{Introduction}

The impressive amount of astrophysical data  which have been accumulated in recent years has depicted a new standard cosmological model according to which the universe is spatially flat and undergoing a phase of accelerated expansion. Strong evidences in favour of this scenario are the Hubble diagram of Type Ia Supernovae (hereafter SNeIa) \cite{SNeIa}, the anisotropy spectrum of the cosmic microwave background radiation (hereafter CMBR) \cite{CMBR} and the matter power spectrum as measured by the clustering properties of the large scale distribution of galaxies \cite{LSS} and the data coming from the Ly$\alpha$ clouds \cite{Lyalpha}. Moreover, the abundance of clusters of galaxies and the gas mass fraction in clusters \cite{fgas} constrain the matter density parameter $\Omega_M \sim 0.3$ thus giving rise to the need for a new component with negative pressure to both close the universe $(\Omega_{tot}\simeq 1)$ and drive its accelerated expansion. This is what is usually referred to as {\it dark energy} whose subtle and elusive nature has opened the doors to an overwhelming flood of papers presenting a great variety of models which try to explain this phenomenon.

The simplest explanation claims for the cosmological constant $\Lambda$ thus leading to the so called $\Lambda$CDM model \cite{Lambda}. Although being the best fit to most of the available astrophysical data \cite{LambdaFit}, the $\Lambda$CDM model is also plagued by many problems on different scales. If interpreted as vacuum energy, $\Lambda$ is up to 120 orders of magnitudes smaller than the predicted value. Furthermore, one should also solve the {\it coincidence problem}, i.e. the nearly equivalence of the matter and $\Lambda$ contribution to the total energy density. As a response to these problems, much interest has been devoted to models with dynamical vacuum energy, dubbed {\it quintessence} \cite{QuintFirst,steinhardt}. These models typically involve scalar fields with a particular class of potentials, allowing the vacuum energy to become dominant only recently (see \cite{QuintRev} for comprehensive reviews). Although quintessence by a scalar field is the most studied candidate for dark energy, it generally does not avoid {\it ad hoc} fine tuning to solve the coincidence problem.

On the other hand, it is worth noting that, despite the broad interest in dark matter and dark energy, their physical properties are still poorly understood at a fundamental level and, indeed, it has never been shown that they are actually two different ingredients apart the observational fact that dark matter furnishes evidences of its existence in clustered structures (as galaxies, clusters of galaxies, etc.) and dark energy is an unclustered component which acts to accelerate the cosmic Hubble flow \cite{moffat,curvgal}. These considerations motivated the great interest recently devoted to  completely different approaches to the quintessence problem. Rather than fine tuning a scalar field potential (which, in several cases, is nothing else but a phenomenological ingredient not motivated by fundamental theories), it is also possible to explain the acceleration of the universe by introducing a cosmic fluid with an equation of state causing it to act like dark matter at high densities and dark energy at low densities. Also this approach is, in some sense, phenomenological but it shows the attractive feature that these models can explain both dark energy and dark matter by a single mechanism (thus automatically solving the coincidence problem) and have therefore been referred to as {\it unified dark energy} (UDE) or {\it unified dark matter} (UDM). Some interesting examples are the generalized Chaplygin gas \cite{Chaplygin}, the tachyonic field \cite{tachyon}, the condensate cosmology \cite{Bruce}, the Hobbit model \cite{Hobbit}, and the scaling dark energy \cite{scaling}. It is worth noting, however, that such models seems to be seriously affected by problems with structure formation \cite{Sandvik} so that some work is still needed before they can be considered as reliable alternatives to dark energy and dark matter.

Actually, there is still a different way to face the  problem of cosmic acceleration. As stressed by Lue et al. \cite{LSS03}, it is possible that the observed acceleration is not the manifestation of another ingredient in the cosmic pie, but rather the first signal of a breakdown of our understanding of the laws of gravitation. From this point of view, it is thus tempting to modify the Friedmann equations to see whether it is possible to fit the astrophysical data with a model comprising only the standard matter. Interesting examples of this kind are the
Cardassian expansion \cite{Cardassian} and the DGP gravity \cite{DGP}.

In this same framework, there is also the attractive possibility to consider the Einstein General Relativity as a particular case of a more fundamental theory. This is the underlying philosophy of what are referred to as $f(R)$ theories of gravity \cite{kerner,capozcurv,MetricRn,review,lnR}. Such theories are a class of the so called Extended Theories of Gravity which have become a sort of paradigm in the study of the gravitational interaction based on corrections and enlargements of the traditional Einstein\,-\,Hilbert scheme. In fact, in the last thirty years, several shortcomings came out in the Einstein scheme and people began to investigate whether General Relativity is the only fundamental theory  capable of explaining the gravitational interaction. Such issues come from cosmology and quantum field theory and essentially are due to the lack of a definitive quantum gravity theory.  Alternative theories of gravity have been pursued in order to attempt, at least, a semi\,-\,classical scheme where General Relativity and its positive results could be recovered. The paradigm consists in adding higher\,-\,order curvature invariants and minimally or non\,-\,minimally coupled scalar fields into dynamics which come out from the effective action of quantum gravity \cite{odintsov}. Other motivations to modify General Relativity come from the issue of a whole recovering of Mach principle \cite{brans} which leads to assume a varying gravitational coupling. All these approaches are not the "{\it full quantum gravity}" but are needed as working schemes toward it. In any case, they are going to furnish consistent and physically reliable results. Furthermore, every unification scheme as Superstrings, Supergravity or Grand Unified Theories, takes into account effective actions where non\,-\,minimal couplings or higher order terms in the curvature invariants come out. Such contributions are due to one\,-\,loop or higher\,-\,loop corrections in the high\,-\,curvature regimes near the full (not yet available) quantum gravity regime \cite{odintsov}. Specifically, this scheme is adopted in order to deal with the quantization on curved spacetimes and the result is that the interactions among quantum scalar fields and background geometry or the gravitational self\,-\,interactions yield corrective terms in the Einstein\,-\,Hilbert Lagrangian \cite{birrell}. Moreover, it has been realized that such corrective terms are inescapable if we want to obtain the effective action of quantum gravity on scales closed to the Planck length \cite{vilkovisky}. Higher\,-\,order terms in curvature invariants (such as $R^{2}$, $R^{\mu\nu} R_{\mu\nu}$,
$R^{\mu\nu\alpha\beta}R_{\mu\nu\alpha\beta}$, $R \,\Box R$, or $R \,\Box^{k}R$) or non\,-\,minimally coupled terms between scalar fields and geometry (such as $\phi^{2}R$) have to be added to the effective Lagrangian of gravitational field when quantum corrections are considered. Furthermore, from a conceptual point of view, there would be no {\it a priori} reason to restrict the gravitational Lagrangian to a linear function of the Ricci scalar $R$, minimally coupled with matter \cite{francaviglia}. Finally, the idea that there are no {\it exact} laws of physics but that the Lagrangians of physical interactions are  {\it stochastic} functions -- with the property that local gauge invariances (i.e. conservation laws) are well approximated in the low energy limit and that physical constants can vary -- has been taken into serious consideration \cite{ottewill}.

Besides fundamental physics motivations, all these  theories have acquired a huge interest in cosmology due to the fact that they {\it naturally} exhibit inflationary behaviours able to overcome the shortcomings of Standard Cosmological Model (based on General Relativity). The related cosmological models seem very realistic and, several times, capable of matching with the observations \cite{starobinsky,la}. Furthermore, it is possible to show that, via conformal transformations, the higher\,-\,order and non\,-\,minimally coupled terms always correspond to Einstein gravity plus one or more than one minimally coupled scalar fields \cite{teyssandier,maeda,wands,gottloeber}. More precisely, every higher\,-\,order term always appears as a contribution of order two in the equations of motion. For example, a term like $R^{2}$ gives fourth order equations \cite{ruzmaikin}, $R \ \Box R$ gives sixth order equations \cite{gottloeber,sixth}, $R \,\Box^{2}R$ gives eighth order equations \cite{eight} and so on. By a conformal transformation, any 2nd\,-\,order of derivation corresponds to a scalar field: for example, fourth\,-\,order gravity gives Einstein plus one scalar field, sixth order gravity gives Einstein plus two scalar fields and so on \cite{gottloeber,schmidt1}. This feature results very interesting if we want to obtain multiple inflationary events since an early stage could select ``very'' large\,-\,scale structures (clusters of galaxies today), while a late stage could select ``small'' large\,-\,scale structures (galaxies today) \cite{sixth}. The philosophy is that each inflationary era is connected with the dynamics of a scalar field. Furthermore, these extended schemes naturally could solve the problem of "graceful exit" bypassing the shortcomings of former inflationary models \cite{la,aclo}.

However, in the weak\,-\,field limit approximation, these theories are expected to reproduce General Relativity which, in any case, is experimentally tested only in this limit \cite{will}. This fact is matter of debate since several relativistic theories {\it do not} reproduce exactly Einstein results in the Newtonian approximation but, in some sense, generalize them. As it was firstly noticed by Stelle \cite{stelle}, a $R^2$ theory gives rise to Yukawa\,-\,like corrections to the Newtonian potential which could have interesting physical consequences. For instance, some authors claim to explain the flat rotation curves of galaxies by using such terms \cite{sanders}. Others \cite{mannheim} have shown that a conformal theory of gravity is nothing else but a fourth\,-\,order theory containing such terms in the Newtonian limit. Besides, indications of an apparent, anomalous, long range acceleration revealed from the data analysis of Pioneer 10/11, Galileo, and Ulysses spacecrafts could be framed in a general theoretical scheme by taking corrections to the Newtonian potential into account \cite{anderson}. In general, any relativistic theory of gravitation can yield corrections to the Newton potential (see for example \cite{schmidt}) which, in the post\,-\,Newtonian (PPN) formalism, could furnish tests for the same theory \cite{will}. Furthermore the newborn {\it gravitational lensing astronomy} \cite{ehlers} is giving rise to additional tests of gravity over small, large, and very large scales which very soon will provide direct measurements for the variation of Newton coupling $G_{N}$ \cite{krauss}, the potential of galaxies, clusters of galaxies \cite{nottale} and several other features of gravitating systems. Such data will be, very likely, capable of confirming or ruling out the physical consistency of General Relativity or of any Extended Theory.

In this paper, we will restrict to $f(R)$ theories and we will face the problem to find the cosmological models (derived from a generic $f(R)$) consistent with data. In this case, the Friedmann equations have to be given away in favour of a modified set of cosmological equations which are obtained by varying a generalized gravity Lagrangian where the scalar curvature $R$ has been replaced by a generic function $f(R)$. The usual General Relativity is recovered in the limit $f(R) = R$, while completely different results may be obtained for other choices of $f(R)$. While in the weak field limit, in particular at Solar System scales, the theory should give the usual Newtonian gravity, at cosmological scales there is an almost complete freedom in the choice of $f(R)$ thus leaving open the way to a wide range of models\footnote{Since, in this class of models, higher order geometrical terms give rise to a quintessential behaviour, it is customary to refer to this scenario also as {\it curvature quintessence} \cite{capozcurv}.}. The key point of $f(R)$ cosmological models is the presence of modified Friedmann
equations obtained by varying the generalized Lagrangian. However, the main problem of this approach lies here since it is not clear how the
variation has to be performed. Actually, once the Robertson\,-\,Walker metric has been assumed, the equations governing the dynamics of the universe are different depending on whether one varies with respect to the metric only or with respect to the metric components and the connections. It is usual to refer to these two possibilities as the {\it metric} and the {\it Palatini approach} respectively. The two methods give the same result only in the case $f(R) = R$, while lead, in general, to significantly different field equations for other choices of the
gravity Lagrangian (see \cite{francaviglia,FFV,ABF04,ACCF} and references therein).

Although much interest has been recently  devoted to the Palatini approach \cite{PalRn}, it is worth noting that $f(R)$ theories were initially investigated in their metric formulation. In particular, in Ref.\,\cite{curvfit}, a toy model with $f(R) = f_0 R^n$ and no matter term was discussed in detail and confronted with the SNeIa Hubble diagram and the age of the universe. Choosing between the Palatini and the metric approach for the $f(R)$ theories is an open problem and a definitive answer is likely far to come. Actually, a significative advantage of the Palatini formulation with respect to  the metric approach to $f(R)$ theories is related to the mathematical simplicity of the dynamical equations that turns out to be of second order. On the other hand, the metric approach leads to a fourth order nonlinear differential equation for the scale factor $a(t)$ that, in general, cannot be solved analytically even for the simplest expressions of $f(R)$. Looking for numerical solutions is a difficult task because of the large uncertainties on the parameters determining the boundary conditions.

It is interesting to wonder whether it is possible to overcome these difficulties with a radically different approach to $f(R)$ theories in the metric formulation. As we will show, the equation governing the dynamic of the universe in higher order theories of gravity may be considered as a linear third order differential equation for $f(R(z))$ as function of the redshift $z$. Moreover, the boundary conditions for $f(R(z))$ are easily set on the basis of physical considerations only and are thus not affected by measurement errors. Since the Ricci curvature scalar $R$ may be also expressed as a function of $z$, it is then immediate to get $f(R)$ and thus determine the gravity Lagrangian. The price to pay is to choose an ansatz for $H(z)$, with $H = \dot{a}/a$ the expansion rate. Such an approach makes it possible to directly reconstruct $f(R)$ from the data since $H(z)$ may be determined, for instance, from the SNeIa Hubble diagram and/or the gas mass fraction in galaxy clusters in a model independent way. Rather than proposing a theory and testing it against the data, we follow here the opposite way around\,: we start from the data and determine which is the $f(R)$ theory that reproduces those data. In this sense, we are facing an {\it inverse problem}.

Although it can be conceived as  an {\it observationally\,-\,based} method, our approach allows to tell something more. As we will see, what is needed to reconstruct $f(R)$ is an expression for $H(z)$, whatever is its underlying motivation. As a consequence, one could adopt for $H(z)$ what is predicted by a given dark energy model (whatever it is) and determine what is the $f(R)$ theory which gives rise to the same dynamics (i.e. the same expansion rate and scale factor). As a consequence, we will show that there is an intrinsic degeneracy among $f(R)$ theories and dark energy models so that these two apparently radically different explanations of cosmic acceleration are reconciled as two different aspects of the same underlying physics.

The outline of  the paper is as follows. In Sect.\,II we derive the cosmological equations from a generic $f(R)$ theory. Sect.\,III is devoted to a detailed explanation of the method to determine $f(R)$ from a given expression for the expansion rate $H$ as function of the redshift $z$. An ansatz for $H(z)$ is suggested starting from astrophysical data or from theoretical motivations. The former case is discussed in Sect.\,IV and makes it possible to reconstruct $f(R)$ directly from the data. The latter possibility instead allows to reconcile $f(R)$ theories with popular dark energy models and is discussed in Sect.\,V. A summary of the results and the conclusions are presented in Sect.\,VI. In the Appendix, we comment on the stability of $f(R)$ theories in the metric formulation.

\section{Curvature quintessence}

Although being a cornerstone of modern physics,  Einstein General Relativity has been definitely {\it experimentally} tested only up to the Solar System scales. As such, it should not come as a surprise if some modifications could be needed on the larger cosmological scales. Furthermore, any attempt of formulating a quantum theory of gravitation claims for a revision of the Einstein theory which could only be attained by generalizing the gravity Lagrangian. Motivated by these considerations, it is worth considering a generic fourth order theory of gravity given by the action\,:

\begin{equation}
{\cal{A}} = \int{d^4x \sqrt{-g} \left [ f(R) + {\cal{L}}_{m} \right ]}
\label{eq: action}
\end{equation}
with $g$ the trace of the metric, $f(R)$ a  generic function of the Ricci scalar curvature $R$ and ${\cal{L}}_m$ the standard matter Lagrangian. Hereafter, we adopt units such that $8 \pi G = c = 1$. The field equations may be obtained by varying with respect to the metric components and can be recast in the expressive form \cite{capozcurv,review}\,:

\begin{equation}
G_{\alpha \beta} = R_{\alpha \beta} -  \frac{1}{2} R g_{\alpha \beta} = T^{(curv)}_{\alpha \beta} + T^{(m)}_{\alpha \beta}
\label{eq: field}
\end{equation}
where we have defined a stress\,-\,energy  tensor for the effective curvature fluid\,:

\begin{eqnarray}
T^{(curv)}_{\alpha \beta} & = & \frac{1}{f'(R)} \left \{ g_{\alpha \beta} \left [ f(R) - R f'(R) \right ] /2 +\right . \nonumber \\
~ & ~ & \nonumber \\
~ & + & \left . f'(R)^{; \mu \nu} \left ( g_{\alpha \mu} g_{\beta \nu} - g_{\alpha \beta} g_{\mu \nu} \right ) \right \}
\label{eq: curvstress}
\end{eqnarray}
and the matter term enters Eq.(\ref{eq: field}) through  the modified stress\,-\,energy tensor\,:

\begin{equation}
T^{(m)}_{\alpha \beta} = \tilde{T}^{(m)}_{\alpha \beta}/f'(R)
\label{eq: mattstress}
\end{equation}
with $\tilde{T}^{(m)}_{\alpha \beta}$ the standard minimally coupled matter stress\,-\,energy tensor. Here and in the following, we denote with a prime the derivative with respect to $R$ and with a dot that with respect to cosmic time $t$. Eqs.(\ref{eq: field})\,-\,(\ref{eq: mattstress}) shows that the effect of geometrical terms may be taken into account by introducing an effective fluid and a coupling of the matter term to the geometry through the function $f'(R)$. Note that the ansatz $f(R) = R + 2 \Lambda$ reproduces the standard General Relativity with a cosmological constant term.

To get the equations governing the dynamics of the universe, we have to choose a metric. The assumed homogeneity and isotropy of the universe motivate the choice of the Robertson\,-\,Walker metric. The modified Friedmann equations are \cite{capozcurv,review}\,:

\begin{equation}
H^2 +\frac{k}{a^2} = \frac{1}{3} \left [ \rho_{curv} + \frac{\rho_m}{f'(R)} \right ] \ , 
\label{eq: fried1}
\end{equation}

\begin{equation}
2 \frac{\ddot{a}}{a} + H^2+\frac{k}{a^2} = - \left ( p_{curv} + p_m \right ) \ , 
\label{eq: fried2}
\end{equation}
where $a(t)$ is the scale factor,  $H = \dot{a}/a$ is the  Hubble parameter, $(\rho_m, p_m)$ are the matter-energy density and pressure and we have defined the same quantities for the effective curvature fluid as\,:

\begin{equation}
\rho_{curv} = \frac{1}{f'(R)} \left \{ \frac{1}{2} \left [ f(R)  - R f'(R) \right ] - 3 H \dot{R} f''(R) \right \} \ ,
\label{eq: rhocurv}
\end{equation}

\begin{equation}
p_{curv} = w_{curv} \rho_{curv}
\label{eq: pcurv}
\end{equation}
with the effective barotropic factor given by\,:

\begin{equation}
w_{curv} = -1 + \frac{\ddot{R} f''(R) + \dot{R} \left [ \dot{R} f'''(R) - H f''(R) \right ]}
{\left [ f(R) - R f'(R) \right ]/2 - 3 H \dot{R} f''(R)} \ .
\label{eq: wcurv}
\end{equation}
Note that the coupling with geometry makes  the Friedmann equations nonlinear in the scale factor $a(t)$ since this quantity enters the definition of the energy density and the pressure of the curvature fluid. Moreover, the Hubble parameter $H$ (and thus the scale factor) determines how the Ricci scalar evolves with time through the following constraint equation\,:

\begin{displaymath}
f''(R) \left [ R + 6 \left ( \frac{\ddot{a}}{a} + H^2 +\frac{k}{a^2}\right ) \right ] = 0 \rightarrow 
\end{displaymath}

\begin{equation}
R = -6 \left ( \dot{H} + 2 H^2 +\frac{k}{a^2}\right ) 
\label{eq: constr}
\end{equation}
where we have used the definition of $H$   and assumed $f''(R) \neq 0$. Note that this latter assumption means that we are indeed considering a generalization of Einstein theory since $f''(R) = 0$ implies $f(R) \propto R$ that is indeed the standard General Relativity. 

There is still another equation that has to be considered. Applying the Bianchi identity to Eq.(\ref{eq: field}), we get a conservation equation for the total energy density\,:

\begin{displaymath}
\dot{\rho}_{tot} + 3 H (\rho_{tot} + p_{tot}) = 0 \ .
\end{displaymath}
Starting from this general scheme, we want to reconstruct the form of $f(R)$ from the Hubble parameter as a function of the redshift $z$. According to the present day observations \cite{CMBR,LSS,fgas}, we can assume a spatially flat universe ($k = 0$) filled with dust matter.

Inserting $\rho_{tot} = \rho_{curv} + \rho_m/f'(R)$  and modelling the matter as dust (i.e. $p_m = 0$), we get\,:

\begin{eqnarray}
\dot{\rho}_{curv} + 3 H (1 + w_{curv}) \rho_{curv} & = & - \frac{1}{f'(R)} (\dot{\rho}_m + 3 H \rho_m) \nonumber \\
~ & ~ & - \rho_m \frac{df'(R)}{dt} \ .
\label{eq: cons}
\end{eqnarray}
Since there is no interaction between the matter and curvature fluid,  we may assume that the matter energy density is conserved so that\,:

\begin{equation}
\rho_m = \Omega_M \rho_{crit} a^{-3} = 3 H_0^2 \Omega_M (1 + z)^3
\label{eq: mattrho}
\end{equation}
with $z = 1/a - 1$ the redshift  (having set $a(t_0) = 1$), $\Omega_M$ the matter density parameter\footnote{Note that $\Omega_M$ is defined in terms of the usual critical energy density, but it is not forced to be unity in a spatially flat universe since the role of $\Omega_{\Lambda}$ is now played by $\Omega_{curv}$ which can be formally derived starting from the curvature quantities.} and hereon quantities labelled with the subscript $0$ refers to present day ($z = 0$) values. Inserting Eq.(\ref{eq: mattrho}) into Eq.(\ref{eq: cons}), we get the following conservation equation for the effective curvature fluid\,:

\begin{eqnarray}
\dot{\rho}_{curv} + 3 H (1 + w_{curv}) \rho_{curv} & = & 3 H_0^2 \Omega_M (1 + z)^3  \nonumber \\
~ & ~ & \times \ \frac{\dot{R} f''(R)}{\left [ f'(R) \right ]^2} \ .
\label{eq: curvcons}
\end{eqnarray}
Actually, the continuity  equation and the two Friedmann equations are not independent since it is possible to show that the dynamics of the universe is completely determined by the two cosmological equations \cite{DK83}. Let us then consider only Eqs.(\ref{eq: fried1}) and (\ref{eq: fried2}). Using the definition of $H$, we may conveniently reduce the system formed by these equations to the following single equation\,:

\begin{displaymath}
\dot{H} + \frac{1}{2} \left [ \frac{\rho_m}{f'(R)} + (1 + w_{curv}) \rho_{curv} \right ] = 0 \ .
\end{displaymath}
Using Eq.(\ref{eq: wcurv})  for $w_{curv}$ and Eq.(\ref{eq: mattrho}) for $\rho_m$, we get\,:

\begin{eqnarray}
\dot{H} & = & -\frac{1}{2 f'(R)} \left \{ 3 H_0^2 \Omega_M (1 + z)^3 + \ddot{R} f''(R)+ \right . \nonumber \\
~ & ~ & \left . + \dot{R} \left [ \dot{R} f'''(R) - H f''(R) \right ] \right \} \ .
\label{eq: presingleeq}
\end{eqnarray}
It is convenient to  change variable from cosmic time $t$ to redshift $z$. To this end, one has simply to use\,:

\begin{displaymath}
\frac{d}{dt} = - (1 + z) H \frac{d}{dz} \ .
\end{displaymath}
With this rule, Eq.(\ref{eq: constr}) can be  written, for $k=0$, as\,:

\begin{equation}
R = -6 \left [ 2 H^2 - (1 + z) H \frac{dH}{dz} \right ] \ .
\label{eq: rvsh}
\end{equation}
Although straightforward  to derive, we find useful  to report the explicit expression of the derivatives of $f(R)$ with respect to $R$ as function of the same derivatives with respect to $z$. It is\,:

\begin{equation}
f'(R) = \left ( \frac{dR}{dz} \right )^{-1} \frac{df}{dz} \ ,
\label{eq: fp}
\end{equation}

\begin{equation}
f''(R) = \left ( \frac{dR}{dz} \right )^{-2} \frac{d^2f}{dz^2} - \left ( \frac{dR}{dz} \right )^{-3} \frac{d^2R}{dz^2} \frac{df}{dz} \ , 
\label{eq: fpp}
\end{equation}

\begin{eqnarray}
f'''(R) & = & \left ( \frac{dR}{dz} \right )^{-3} \frac{d^3f}{dz^3}
+ 3 \left ( \frac{dR}{dz} \right )^{-5} \left ( \frac{d^2R}{dz^2} \right )^2 \frac{df}{dz} +\nonumber \\
~ & - & \left ( \frac{dR}{dz} \right )^{-4} \left ( 3 \frac{d^2R}{dz^2} \frac{d^2f}{dz^2} + \frac{d^3R}{dz^3} \frac{df}{dz} \right ) \ .
\label{eq: f3p}
\end{eqnarray}
Using Eqs.(\ref{eq: rvsh})\,-\,(\ref{eq: f3p}), it is possible to rewrite Eq.(\ref{eq: presingleeq}) as a differential equation containing only the unknown functions $H(z)$ and $f(z)$ and their derivatives with respect to $z$. Note that we are using the abuse of notation $f(R(z)) = f(z)$. Some simple but lenghty algebra allows to rewrite Eq.(\ref{eq: presingleeq}) as\,:

\begin{equation}
{\cal{H}}_3(z) \frac{d^3f}{dz^3} + {\cal{H}}_2(z) \frac{d^2f}{dz^2} + {\cal{H}}_1(z) \frac{df}{dz} = - 3 H_0^2 \Omega_M (1 + z)^3
\label{eq: singleeq}
\end{equation}
with\,:

\begin{eqnarray}
{\cal{H}}_1 & = & \dot{R}^2 \left ( \frac{dR}{dz} \right )^{-4} \left [ 3 \left ( \frac{dR}{dz} \right )^{-1}
\left ( \frac{d^2R}{dz^2} \right )^{2} - \frac{d^3R}{dz^3} \right ]+ \nonumber \\
~ & ~ & - \left ( \ddot{R} - \dot{R} H \right ) \left ( \frac{dR}{dz} \right )^{-3} \frac{d^2R}{dz^2} +\nonumber \\
~ & ~ & -2 (1 + z) H \frac{dH}{dz} \left ( \frac{dR}{dz} \right )^{-1} \ ,
\label{eq: h1}
\end{eqnarray}

\begin{equation}
{\cal{H}}_2 = \left ( \ddot{R} - \dot{R} H \right ) \left ( \frac{dR}{dz} \right )^{-2} - 3 \dot{R}^2 \left ( \frac{dR}{dz} \right )^{-4} 
\frac{d^2R}{dz^2} \ ,
\label{eq: h2}
\end{equation}

\begin{equation}
{\cal{H}}_3 = \dot{R}^2 \left ( \frac{dR}{dz} \right )^{-3} \ .
\label{eq: h3}
\end{equation}
The derivatives of $R$ may be evaluated from Eq.(\ref{eq: rvsh}). For instance, differentiating with respect to $z$, we get\,:

\begin{equation}
\frac{dR}{dz} = -6 \left \{ -(1 + z) \left ( \frac{dH}{dz} \right )^2 + H \left [ 3 \frac{dH}{dz} - (1 + z) \frac{d^2H}{dz^2} \right ] \right \}
\label{eq: drdz}
\end{equation}
and so on for higher order derivatives.  For completeness, we also report the following useful relations\,:

\begin{equation}
\dot{R} = - (1 + z) H \frac{dR}{dz} \ ,
\label{eq: dotR}
\end{equation}

\begin{eqnarray}
\ddot{R} - \dot{R} H & = & 6 (1 + z) H^2 \left \{ 3 (1 + z)^2 \frac{dH}{dz} \frac{d^2H}{dz^2} +\right . \nonumber \\
~ & ~ & + \left . H \left [ (1 + z)^2 \frac{d^3H}{dz^3}  - 6 \frac{dH}{dz} \right ] \right \} \ .
\label{eq: ddotR}
\end{eqnarray}
Inserting Eqs.(\ref{eq: drdz})\,-\,(\ref{eq: ddotR}) into Eq.(\ref{eq: singleeq}), we get a cumbersome differential equation which we do not report here for sake of shortness. Our task is now to solve this equation in order to obtain a form of $f(R(z))$ from the Hubble parameter $H(z)$.

\section{Solving with respect to $f(z)$}

In the usual metric approach  to $f(R)$ theories in cosmology, one chooses an expression for $f(R)$, uses Eq.(\ref{eq: rvsh}) to replace $R$ with $H$ and finally get Eq.(\ref{eq: presingleeq}) as a differential equation for the scale factor. Unfortunately, here the game is over. Indeed, Eq.(\ref{eq: singleeq}) is a fourth order nonlinear differential equation for $a(t)$ which is difficult to solve analytically also for the simplest models of $f(R)$ such as power law or logarithmic. To overcome this problem, one could impose an analytical ansatz for $a(t)$ and look whether and under which conditions the chosen expression solves Eq.(\ref{eq: presingleeq}). Actually, this approach is quite unsatisfactory since there are no hints which may suggest a possible expression for $a(t)$ so that one has to perform a blind search with the risk of wasting time without ending with any successful result. Moreover, even if such a solution were found, it should be of limited applicability so that drawing any conclusion on the viability of a given $f(R)$ theory on the basis of this particular solution is dangerous. A possible way to overcome this problem is to resort to numerical solutions. However, also this approach is plagued by its own problems. Actually, numerically solving Eq.(\ref{eq: presingleeq}) demands for the knowledge of boundary conditions. Since the equation is of fourth order, one should give the present day values of the scale factor and its derivatives up to the third order. From an observational point of view, this means that one should set the values of $a_0$, the Hubble constant $H_0$, the deceleration parameter $q_0$ and the jerk parameter $j_0$ \cite{Visser} or one should take into account another set of parameters (see e.g. \cite{corasaniti} for a discussion). While in a flat universe we may set $a_0 = 1$ and $H_0$ is reasonably well constrained, setting $q_0$ and $j_0$ is quite complicated\footnote{Note that the boundary conditions may also be set using the statefinder parameter $r_0$ \cite{sf} instead of the jerk $j_0$.}. It is worth stressing that $q_0$ and $j_0$ should be estimated in a model independent way. To this end, one may Taylor expand the scale factor up to the fourth order (or to higher orders) and fit the corresponding luminosity distance to the SNeIa Hubble diagram \cite{series}. While this is technically possible, the results are affected by quite large errors (especially for $j_0$) and, moreover, are somewhat dependent on the order to which the series is truncated\footnote{To be more precise, the best fit values of $q_0$ and $j_0$ are almost unaltered, but the uncertainties get larger as the order of the series expansion increases.}. As a consequence, one should numerically solve Eq.(\ref{eq: presingleeq}) for a quite large set of boundary conditions thus leading to a so great uncertainty on the reconstructed $a(t)$ that it remains practically undetermined.

Given this situation, it is worth asking whether a radically different approach is possible. To this end, let us first observe that what is indeed determined from the data is the Hubble parameter $H(z)$ since this is the quantity entering the definition of both the luminosity and angular diameter distances which are tested by the SNeIa Hubble diagram and the data on the gas mass fraction in galaxy clusters respectively. Actually, this is what is usually done when matching any dark energy model to this kind of astrophysical data. Given a background cosmological model, one first compute the corresponding $H(z; {\bf p})$ and then constrains the model parameters ${\bf p}$ by comparing with the data. Moreover, $H(z)$ could also be directly recovered from the luminosity distance in a model independent way \cite{HzDl,DlPol} although this procedure is still affected by large errors. Motivated by these considerations, we try to go backward from the data to $f(R)$ by assuming an ansatz for the Hubble parameter $H(z)$ rather than for $f(R)$ itself. Inserting an ansatz for $H(z)$ into Eqs.(\ref{eq: h1})\,-\,(\ref{eq: ddotR}), Eq.(\ref{eq: singleeq}) can be seen as a third order differential equation for the function $f(z)$. Since $R(z)$ may be straightforwardly evaluated from Eq.(\ref{eq: rvsh}) for a given $H(z)$, we may then obtain $f(R)$ by eliminating (numerically) the redshift $z$ from the solution $f(z)$ of Eq.(\ref{eq: singleeq}). Note that considering $f(z)$ as unknown has two immediate advantages. First, the equation is of third rather than fourth order so that the numerical solution is easier to find. Second, the equation is linear in $f(z)$ and its derivatives so that, for given boundary conditions, the solution exists and is unique.

To numerically solve the differential equation  for $f(z)$, one has to set the boundary conditions, i.e. the values of $f$ and its first and second derivatives with respect to $z$ evaluated at $z = 0$. Here lies another advantage of this approach\,: the boundary conditions may be chosen on the basis of physical considerations only. To this end, let us first remind that it has been shown that, in order to not contradict Solar System tests of gravity, a whatever $f(R)$ theory must fulfill the condition $f''(R_0) = 0$ \cite{newtlimitok}. Using Eq.(\ref{eq: fpp}), we thus get the constraint\,:

\begin{displaymath}
f''(R_0) = \left [ \left ( \frac{dR}{dz} \right )^{-2} \frac{d^2f}{dz^2} -  \left ( \frac{dR}{dz} \right )^{-3}
\frac{d^2R}{dz^2} \frac{df}{dz} \right ]_{z = 0} \ .
\end{displaymath}
A second condition may be obtained considering Eq.(\ref{eq: fried1}).   Let us rewrite it introducing back the coupling factor $8 \pi G$ and $k=0$. It is\,:

\begin{displaymath}
H^2 = \frac{8 \pi G}{3} \left [ \rho_{curv} + \frac{\rho_m}{f'(R)}
\right ]\ .
\end{displaymath}
This equation shows that the coupling of the matter with geometry through the function $f'(R)$ is equivalent to redefine the Newton gravitational constant $G$ as $G/f'(R)$ so that we get the well known result that in $f(R)$ theories $G$ is redshift (and hence time) dependent. Indeed, at $z = 0$, the effective gravitational constant $G/f'(R_0)$ must be equal to $G$ so that we get\,:

\begin{displaymath}
f'(R_0) = 1 \rightarrow \left [ \left ( \frac{dR}{dz} \right )^{-1} \frac{df}{dz} \right ]_{z = 0} = 1 \ .
\end{displaymath}
Combinining the constraints on $f'(R_0)$ and $f''(R_0)$,  we get the following boundary conditions\,:

\begin{equation}
\left ( \frac{df}{dz} \right )_{z = 0} = \left ( \frac{dR}{dz} \right )_{z = 0} \ ,
\label{eq: fpzero}
\end{equation}

\begin{equation}
\left ( \frac{d^2f}{dz^2} \right )_{z = 0} = \left ( \frac{d^2R}{dz^2} \right )_{z = 0} \ .
\label{eq: fppzero}
\end{equation}
A comment  is in order here. We have derived the present day values of $df/dz$ and $d^2f/dz^2$ by imposing the consistency of the reconstructed $f(R)$ theory with {\it local} Solar System test. One could wonder whether tests on local scales could be used to set the boundary conditions for a cosmological problem. It is easy to see that this is indeed meaningful. Actually, the isotropy and homogeneity of the universe ensure that the present day value of a whatever cosmological quantity does not depend on where the observer is. As a consequence, an hypothetical observer living in the Andromeda galaxy and testing gravity in his planetary system should get the same results. As such, the present day values of $df/dz$ and $d^2f/dz^2$ adopted by this hypothetical observer are the same as those we have used based on our Solar System experiments. Therefore, there is no systematic error induced by our method of setting the boundary conditions.

Finally, to set $f(z = 0)$, let us evaluate the  present day value of  $\rho_{curv}$ using the definition (\ref{eq: rhocurv}) and Eqs.(\ref{eq: fpzero}) and (\ref{eq: fppzero}). We get\,:

\begin{displaymath}
\rho_{curv}(z = 0) = \frac{f(R_0) - R_0}{2} \ .
\end{displaymath}
Evaluating Eq.(\ref{eq: fried1}) at $z = 0$ and solving  with respect to $f(R_0)$, we finally get\,:

\begin{equation}
f(z = 0) = f(R_0) = 6 H_0^2 (1 - \Omega_M) + R_0 \ .
\label{eq: fzero}
\end{equation}
Summarizing, to reconstruct $f(R)$ from the data,  we adopt the procedure schematically sketched below.

\begin{enumerate}

\item{Assume an expression for $H(z)$ and determine the model parameters from fitting to the data.}

\item{Compute $R(z)$ from Eq.(\ref{eq: rvsh}).}

\item{Use Eqs.(\ref{eq: h1})\,-\,(\ref{eq: ddotR}) to write Eq.(\ref{eq: singleeq}) for $f(z)$.}

\item{Solve numerically the above equation with the boundary conditions given by Eqs.(\ref{eq: fpzero})\,-\,(\ref{eq: fzero}).}

\item{With the aid of the derived expressions for $R(z)$ and $f(z)$, eliminate $z$ from $f(z)$ and finally get $f(R)$.}

\end{enumerate}
This quick pipeline\footnote{A fast and efficient code implementing this procedure,  written for the {\it Mathematica 4.1} software, is available on request.} makes it possible to obtain an expression for $f(R)$ that is observationally well founded since the $f(R)$ theory thus reconstructed fits the same dataset used to determine the parameters entering $H(z)$. It is worth stressing that such a result has been obtained without the need to solve Eq.(\ref{eq: presingleeq}) with respect to the scale factor $a(t)$. Moreover, any arbitraryness in the choice of an expression for $f(R)$ has been removed.

\section{Determining $f(R)$ from the data}

A key ingredient  in the procedure outlined above is  the choice of an analytical expression for the dependence of the Hubble parameter on the redshift. Rather than choosing a somewhat motivated ansatz, one should resort to an empirical determination of $H(z)$ from the data. When used as input for our pipeline, this function makes it possible to reconstruct $f(R)$ directly from the data thus avoiding any systematic bias or theoretical prejudice. We will consider here two different approaches to implement an {\it observations\,-\,based} determination of $f(R)$.

\subsection{$f(R)$ from $D_L(z)$}

Fitting to the SNeIa  Hubble diagram is nowaday a standard  tool to investigate the viability of a given cosmological model. The essence of the method relies on the well known relation\,:

\begin{equation}
\mu(z) = 5 \log{D_L(z)} + 25
\label{eq: dm}
\end{equation}
being $\mu$ the distance modulus of an object at redshift $z$ and $D_L$ its luminosity distance (in Mpc) which is determined by the Hubble parameter as follows\footnote{Remember that we are using units such that $c = 1$.}\,:

\begin{equation}
D_L(z) = (1 + z) \int_0^z{\frac{d\zeta}{H(\zeta)}} \ .
\label{eq: dl}
\end{equation}
This equation may be inverted to give \cite{HzDl}\,:

\begin{equation}
H(z) = \left \{ \frac{d}{dz} \left [ \frac{D_L(z)}{1 + z} \right ] \right \}^{-1}
\label{eq: hzdl}
\end{equation}
so that $H(z)$ may be determined from $D_L(z)$.  Because of Eq.(\ref{eq: dm}), the luminosity distance may be empirically determined by measuring the distance modulus for a class of standard candles thus allowing a direct reconstruction of $H(z)$ from the data in a model independent way.

It is worth wondering how much accurate the  determination  of $\mu(z)$ must be in order to efficiently recover $H(z)$ and thus $f(R)$. A quantitative answer requires detailed simulations and it is outside the scope of the present paper. Actually, we may get a qualitative understanding of the problem by considering the error on $R(z)$ due to the measurement uncertainties on $\mu(z)$. To this end, let us first insert Eq.(\ref{eq: hzdl}) into Eq.(\ref{eq: rvsh}) to get\,:

\begin{equation}
R(z) = \frac{6 (1 + z)^6 D''_L(z)}{\left [ D_L(z) - (1 + z) D'_L(z) \right ]^3}
\label{eq: rvsdl}
\end{equation}
where, only in this section, the prime denotes the  derivative with respect to $z$. If the errors on $D_L$, $D'_L$ and $D''_L$ are Gaussian  distributed, we can determine the uncertainty $\sigma_R$ on $R$ by the usual rules. This is likely to be not the case. Nonetheless, propagating the errors on $D_L$, $D'_L$ and $D''_L$ should give us an order of magnitude to estimate $\sigma_R$ that is enough for our aims. To this end, let us denote with $(\delta_0, \delta_1, \delta_2)$ the quantities $(D_L, D'_L, D''_L)$ respectively and let $(\sigma_0, \sigma_1, \sigma_2)$ the corresponding uncertainties. A naive estimate of $\sigma_R$ may be obtained as\,:

\begin{displaymath}
\sigma_R = \sqrt{\left | \frac{\partial R}{\partial \delta_0} \right |^2 \sigma_0^2
+ \left | \frac{\partial R}{\partial \delta_1} \right |^2 \sigma_1^2
+ \left | \frac{\partial R}{\partial \delta_2} \right |^2 \sigma_2^2} \ .
\end{displaymath}
Inserting Eq.(\ref{eq: rvsdl}) into the above relation, we get\,:

\begin{equation}
\sigma_R = \left | \frac{6 (1 + z)^6}{\left [ \delta_0 - (1 + z) \delta_1 \right ]^3} \right | \times \Delta_R
\label{eq: sigr}
\end{equation}
with\,:

\begin{equation}
\Delta_R = \sqrt{\left | \frac{3 \delta_2 \ \sigma_0}{(1 + z) \delta_1 - \delta_0} \right |^2 +
\left | \frac{3 (1 + z) \delta_2 \ \sigma_1}{\delta_0 - (1 + z) \delta_1} \right |^2 + \sigma_2^2}
\label{eq: deltar}
\end{equation}
Inverting Eq.(\ref{eq: dm}), we may express $(\delta_0, \delta_1, \delta_2)$  as  functions of the distance modulus $\mu(z)$ and its derivatives up to the second order. Following the same procedure for propagating the errors, we get\,:

\begin{equation}
\sigma_0 = k \ \delta_0 \ \sigma_{\mu} \ ,
\label{eq: sigzero}
\end{equation}

\begin{equation}
\sigma_1 = k \sqrt{\mu_1^2 \sigma_0^2 + \delta_0^2 \sigma_{\mu_1}^2} \ ,
\label{eq: siguno}
\end{equation}

\begin{equation}
\sigma_2 = k \sqrt{(k \mu_1^2 + \mu_2)^2 \sigma_0^2 + k (k \sigma_{\mu_1}^2 + \sigma_{\mu_2}^2)^2 \delta_0^2} \ ,
\label{eq: sigdue}
\end{equation}
having set $k = \ln{(10)}/5$, $\mu_1 = d\mu/dz$, $\mu_2 = d^2\mu/dz^2$   and denoted with $(\sigma_{\mu}, \sigma_{\mu_1}, \sigma_{\mu_2})$ the measurement uncertainties on $(\mu, \mu_1, \mu_2)$ respectively. Note that Eqs.(\ref{eq: sigzero})\,-\,(\ref{eq: sigdue}) clearly show that the errors on the luminosity distance and its derivatives are correlated, while Eq.(\ref{eq: sigr}) has been obtained considering the errors as uncorrelated. As a consequence, Eq.(\ref{eq: sigr}) underestimates $\sigma_R$. Nonetheless, some interesting conclusions on $\sigma_R$ may be drawn. First, we note that $\sigma_R \propto (1 + z)^6$ so that the error on the reconstructed Ricci scalar quickly increases with the redshift. This conclusion is further strengthned when considering that $(\delta_0, \delta_1, \delta_2)$ become larger and larger with $D_L$. Although a detailed investigation (with the aid of simulations) is needed, one should argue that the distance modulus must be measured with a tiny uncertainty in order to reduce as more as possible $\sigma_R$. Moreover, Eqs.(\ref{eq: rvsdl}) and (\ref{eq: sigr}) show that both $R$ and $\sigma_R$ depend on the derivatives up to the second order of the distance modulus with respect to the redshift. Actually, observationally constraining $d\mu/dz$ and $d^2\mu/dz^2$ is a quite complicated task likely demanding for very large samples of SNeIa. Since reconstructing $f(R)$ needs first a determination of $R(z)$, we may thus conclude that a completely model independent determination of $f(R)$ from the luminosity distance data is plagued by too large uncertainties and is far to come.

A possible way to escape these problems is to resort to some parametrized expression of $D_L(z)$ so that one has only to constrain a limited set of quantities rather than recover a function (i.e., determine an infinite number of unknowns). To this end, a series expansion \cite{DlPol} or a more versatile fitting function \cite{DlFit} have been proposed. Inserting the expression for $D_L$ into Eqs.(\ref{eq: rvsdl}) and (\ref{eq: sigr}), one could evaluate both $R$ and $\sigma_R$ as function of the parameters assigning the luminosity distance so that the problem of reconstructing $R(z)$ reduces to the determination of these parameters. Similar considerations also hold for $f(z)$. Calibrated simulations are needed to investigate the viability of this promising approach.

\subsection{Polynomial fit to the dark energy density}

The problem to recover $f(R)$ from the luminosity distance in a model independent way is partially related to the need of determining from the data not only $D_L(z)$, but its derivatives too. To overcome this difficulty, one should resort to a less ambitious program choosing a parametric ansatz for $H(z)$ which should be as general as possible. A useful expression is \cite{HzPol}\,:

\begin{equation}
H(z) = H_0 \sqrt{\Omega_M x^3 + A_1 + A_2 x + A_3 x^2}
\label{eq: hzpol}
\end{equation}
with $x = 1 + z$ and $(A_1, A_2, A_3)$ parameters to  be determined from the data. Note that the case $(A_1, A_2, A_3) = (0, 0, 0)$ corresponds to a flat matter only universe, while the case $A_1 \ne 0$ and $(A_2, A_3) = (0, 0)$ gives $H(z)$ for the $\Lambda$CDM concordance model. In general, Eq.(\ref{eq: hzpol}) refers to a universe containing a matter term and a second component whose energy density is approximated by a second order polynomial fit.

This expression has the remarkable property of simplicity. Actually, by inserting Eq.(\ref{eq: hzpol}) into Eq.(\ref{eq: rvsdl}), we get\,:

\begin{equation}
R = -3 H_0^2 (4 A_1 + 3 A_2 + 2 A_3 + \Omega_M) \ ,
\label{eq: rpol}
\end{equation}
while the boundary conditions turn out to be given as\,:

\begin{equation}
f(z = 0) = -3 H_0^2 (4 A_1 + 3 A_2 + 2 A_3 + 3 \Omega_M - 2) \ ,
\label{eq: fzpol}
\end{equation}

\begin{equation}
df/dz |_{z = 0} = -3 H_0^2 (3 A_2 + 4 A_3 + 3 \Omega_M) \ ,
\label{eq: fppol}
\end{equation}

\begin{equation}
d^2f/dz^2 |_{z = 0} = -3 H_0^2 (2 A_3 + 3 \Omega_M) \ .
\label{eq: f2ppol}
\end{equation}
Denoting with $\sigma_i$ the uncertainty on $A_i$ and with $(\sigma_H, \sigma_M)$  those on $H_0$ and $\Omega_M$ respectively, a naive estimate of the error on $R$ is given by\,:

\begin{equation}
\sigma_R = 3 H_0^2 \sqrt{(12/H_0^2) \sigma_H^2 + 16 \sigma_1^2 + 9 \sigma_2^2 + 4 \sigma_3^2 + \sigma_M^2}
\label{eq: sigrpol}
\end{equation}
having assumed that the errors are statistical  and uncorrelated. It is worth noting that $\sigma_R$ turn out to be independent of the redshift $z$ so that a reliable reconstruction of the evolution of the scalar curvature only demands for lowering the errors on the parameters $(H_0, \Omega_M, A_1, A_2, A_3)$. This could be achieved by increasing the sample of SNeIa or reducing the measurement and systematic errors on the distance modulus or by a judicious use of priors on $(H_0, \Omega_M)$.

Having reconstructed $R(z)$, one has only to  determine $f(z)$. Inserting Eq.(\ref{eq: hzpol}) into Eq.(\ref{eq: singleeq}), we get\,:

\begin{equation}
{\cal{P}}_3(z) \frac{d^3f}{dz^3} + {\cal{P}}_2(z) \frac{d^2f}{dz^2}
+ {\cal{P}}_1(z) \frac{df}{dz} = -3 H_0^2 \Omega_M (1 + z)^3
\label{eq: singpol}
\end{equation}
where ${\cal{P}}_i$ are cumbersome functions of the redshift $z$ and the parameters $(\Omega_M, A_1, A_2, A_3)$ that may be computed inserting Eq.(\ref{eq: hzpol}) into Eqs.(\ref{eq: h1})\,-\,(\ref{eq: ddotR}). We do not report them for sake of shortness. Eq.(\ref{eq: singpol}) may be straightforwardly solved numerically using the boundary conditions (\ref{eq: fzpol})\,-\,(\ref{eq: f2ppol}) so that $f(z)$ may be obtained and coupled with the above reconstructed $R(z)$ to finally determine $f(R)$ from the data. Estimating what is the error on $f(R)$ is quite complicated and it is likely that a numerical analysis is needed.

\section{$f(R)$ theories and dark energy models}

As discussed in the previous section, $H(z)$ may be directly reconstructed from the data on the luminosity distance. Although in principle possible, this method is nowaday still not feasible since it is affected by quite large errors. Furthermore, $H(z)$ is thus recovered only on the redshift range probed by the data used so that a potentially dangerous extrapolation is needed to go to higher $z$. This problem forces us to adopt for $H(z)$ a theoretically rather than observationally motivated ansatz. To this end, it is worth referring to the literature where the accelerated expansion of the universe is usually explained proposing a wide variety of dark energy models. Although different in their physical background, all these models share the property of well fitting the same dataset so that, from this point of view, we could adopt for $H(z)$ the expression corresponding to one of these models. This choice also allows us to stress one interesting point. Since the procedure above sketched makes it possible to recover $f(R)$ from $H(z)$, we are able to construct an $f(R)$ theory of gravity which gives the same dynamics (i.e., the same expansion rate and scale factor) of a whatever dark energy model. We will show this explicitly for two popular dark energy models, namely quintessence with constant equation of state referred to as {\it quiessence} and the Chaplygin gas.

As a preliminary remark, it is worth  discussing what  are the dimensions of the different quantities involved in the units (with $8 \pi G = c = 1$) we are adopting. To this end, let us first consider Eq.(\ref{eq: constr}). Since $H$ is expressed in $s^{-1}$, the Ricci scalar turns out to be measured in $s^{-2}$. From Eq.(\ref{eq: fried1}) and the consideration that $\rho_{curv}$ and $\rho_m$ have the same dimensions, we conclude that $f'(R) = df/dR$ is dimensionless and hence $f$ has the same dimension as $R$. Finally, the energy density is expressed in $s^{-2}$ as can be inferred from the expression of the critical density that is $\rho_{crit} = 3 H_0^2$ in these units. It is straightforward to check that this is consistent with what is obtained for $\rho_{curv}$ from Eq.(\ref{eq: rhocurv}). Hereon, we measure time in units of $1/H_0$ so that $H_0 = 1$ and all the quantities we are interested in, namely $R$ and $f(R)$, are dimensionless that is a useful feature when dealing with numerically solving differential equations.

\subsection{Quiessence from $f(R)$}

As a first straightforward application, let us consider the ansatz\,:

\begin{equation}
H(z) = H_0 \sqrt{\Omega_M (1 + z)^3 + \Omega_X (1 + z)^{3 (1 + w)}}
\label{eq: hqcdm}
\end{equation}
with $\Omega_X = (1 - \Omega_M)$ and $w$  a constant parameter. Eq.(\ref{eq: hqcdm}) gives the Hubble parameter for the so called quiessence models (or QCDM) where the acceleration of the universe is due to a negative pressure fluid with constant barotropic factor $w$. This is the easiest generalization of the cosmological constant which is obtained for $w = -1$. Quiessence has been successfully tested against the SNeIa Hubble diagram and the CMBR anisotropy spectrum that has made it possible to severely constraint the barotropic factor $w$ \cite{EstW}. It is interesting to note that these constraints extend into the region $w < -1$ so that models violating the weak energy condition are allowed. This class of models, dubbed {\it phantom models}, are affected by serious problems with the growth of perturbations and are therefore worrisome.

\begin{figure}
\centering \resizebox{8.5cm}{!}{\includegraphics{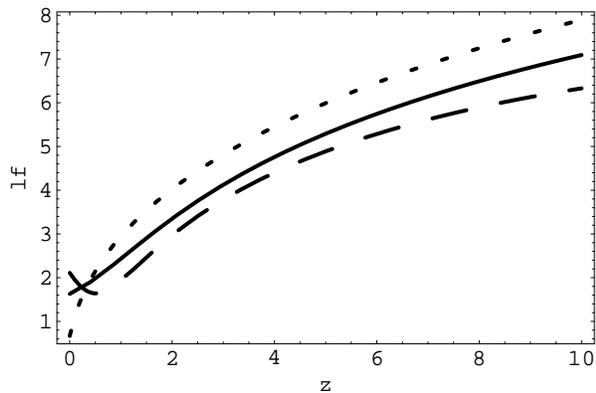}}
\caption{Reconstructed $f(z)$ for quiessence models with $\Omega_M = 0.3$   and three different values of the dark energy barotropic factor, namely $w = -0.5$ (short dashed), $w = -1$ (solid) and $w = -1.5$ (long dashed). We report $lf = \ln{(-f)}$ rather than $f$ to better show the results.} 
\label{fig: fvszqcdm}
\end{figure}

\begin{figure}
\centering \resizebox{8.5cm}{!}{\includegraphics{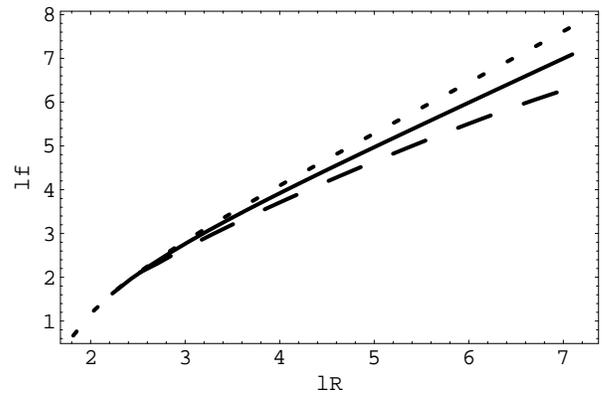}}
\caption{Reconstructed $f(R)$ for quiessence models with $\Omega_M = 0.3$   and three different values of the dark energy barotropic factor, namely $w = -0.5$ (short dashed), $w = -1$ (solid) and $w = -1.5$ (long dashed). We report $lf = \ln{(-f)}$ as function of $lR = \ln{(-R)}$ rather than $f(R)$ to better show the results.}
\label{fig: fvsrqcdm}
\end{figure}

Inserting Eq.(\ref{eq: hqcdm}) into Eq.(\ref{eq: rvsh}), we get\,:

\begin{equation}
R = - 3 H_0^2 \left [ \Omega_M (1 + z)^3 + \Omega_X (1 - 3 w) (1 + z)^{3 (1 + w)} \right ] \ .
\label{eq: rvszqcdm}
\end{equation}
Note that $R$ is always negative as a consequence of the signature $\{+, -, -, -\}$ adopted.  If we had used the opposite signature, Eq.(\ref{eq: rvszqcdm}) is the same, but with an overall positive sign. The ansatz in Eq.(\ref{eq: hqcdm}) leads to the following equation for $f(z)$\,:

\begin{equation}
{\cal{Q}}_3(z) \frac{d^3f}{dz^3} + {\cal{Q}}_2(z) \frac{d^2f}{dz^2} + {\cal{Q}}_1(z) \frac{df}{dz} = -3 H_0^2 \Omega_M (1 + z)^3
\label{eq: eqqcdm}
\end{equation}
where ${\cal{Q}}_i(z)$ may be evaluated by inserting Eq.(\ref{eq: hqcdm}) into Eqs.(\ref{eq: h1})\,-\,(\ref{eq: ddotR}). The resulting cumbersome functions of the redshift $z$ and the model parameters $(\Omega_M, w)$ are not reported here for sake of shortness. Eq.(\ref{eq: eqqcdm}) may be easily solved numerically using the boundary conditions (\ref{eq: fpzero})\,-\,(\ref{eq: fzero}). The result is shown in Fig.\,\ref{fig: fvszqcdm} for models with $\Omega_M = 0.3$ and three different choices of the barotropic factor $w$. Note that we have plotted $-f(z)$ rather than $f(z)$ since $f$ turns out to be negative because of the signature adopted. Moreover, we use a logarithmic scale to better show the results. For $z < 0.5$ the three curves show a different behaviour, while, for higher $z$, the shape of $f(z)$ is unaffected by $w$ which only acts as a scaling parameter. This is also true for $\Omega_M$\,: changing this parameter only shifts up or down the curves plotted in Fig.\,\ref{fig: fvszqcdm}. Inverting numerically Eq.(\ref{eq: rvszqcdm}), we may obtain $z = z(R)$ and finally get $f(R)$ shown in Fig.\,\ref{fig: fvsrqcdm} for the same models considered above. It turns out that $f(R)$ is the same for different models for low values of $R$ and hence of $z$. This is a consequence of the well known degeneracy among different quiessence models at low $z$ that, in the standard analysis, leads to large uncertainties on $w$. Because of this degeneracy, models with different barotropic factors may not be discriminated and are therefore dynamically equivalent. This is reflected in the shape of the reconstructed $f(R)$ that is almost $w$\,-\,independent in this redshift range.

Fig.\,\ref{fig: fvsrqcdm} may suggest that $\ln{(-f)}$ is a quasi\,-\,linear function of $\ln{(-R)}$ that is not the case. Actually, we have checked that the following empirical function

\begin{equation}
\ln{(-f)} = l_1 \left [ \ln{(-R)} \right ]^{l_2} \left [ 1 + \ln{(-R)} \right ]^{l_3} + l_4
\label{eq: fit}
\end{equation}
approximates very well the numerical  solution provided that the parameters $(l_1, l_2, l_3, l_4)$ are suitably chosen for a given value of $w$. For instance, for $w = -1$ (the cosmological constant\footnote{The case of the cosmological constant may be solved analytically giving $f(R) = R + 2 \Lambda$ which is quite different from Eq.(\ref{eq: fit}). Actually, Eq.(\ref{eq: fit}) is only an approximating function chosen because of its versatilty. Indeed, for $w = -1$, the fitting parameters renders the approximating function as similar as possible to the exact expression.}) it is\,:

\begin{displaymath}
(l_1, l_2, l_3, l_4) = (2.6693, 0.5950, 0.0719, -3.0099) \ .
\end{displaymath}
The error in this case is of the order  $\sim 2 - 10\%$ for $z < 1$,  but remains smaller than $\sim 4\%$ up to $z = 10$. We have checked that Eq.(\ref{eq: fit}) works very well also for other values of $w$ in the range $(-1.6, -0.6)$ with an error\footnote{Tables with the values of the parameters and the approximation error for different $w$ and $z$ are available on request.} depending on $z$ and $w$, but always smaller than $\sim 10\%$.

Some comments are in order as final remarks.  First, remember that we are using units with $8 \pi G = c = 1$ and $1/H_0 = 1$. As a consequence, the values of the fitting parameters should be adjusted if physical units are used. However, this does not change the shape of the approximating function since the numerical solution does not depend on the adopted units. Let us also stress that Eq.(\ref{eq: fit}) is only a fitting function tested over the large, but still limited redshift range $(0, 10)$. Extrapolation to higher $z$ may introduce large errors and should therefore be avoided.

\subsection{Chaplygin gas as $f(R)$\,-\,model}

Recently, much attention  has been devoted to models where a single fluid with an exotic equation of state accounts for both dark matter and dark energy. Usually referred to as {\it unified dark energy models}, such models have the nice feature to solve two problems in a single step and have thus attracted a lot of interest. As a prototype example, we consider the generalized Chaplygin gas (GCG) whose equation of state is \cite{Chaplygin}\,:

\begin{equation}
p = -A/\rho^{\alpha}
\label{eq: pch}
\end{equation}
with $A$ and $\alpha$ positive parameters to be  determined from the data.  Assuming a spatially flat universe, the Hubble parameter is\,:

\begin{equation}
H(z) = H_0 \left [ A_s + (1 - A_s) (1 + z)^{3 (1 + \alpha)} \right ]^{\frac{1}{2 (1 + \alpha)}}
\label{eq: hcg}
\end{equation}
with $A_s = A/\rho(z = 0)$. Note that $A_s = 1$  reduces to a universe with the cosmological constant as unique component, while $A_s = 0$ describes a matter dominated universe. Motivated by this consideration, we will only take into account models with $A_s$ in the range $(0, 1)$ although, in principle, nothing prevents $A_s$ to be larger than 1.

\begin{figure}
\centering \resizebox{8.5cm}{!}{\includegraphics{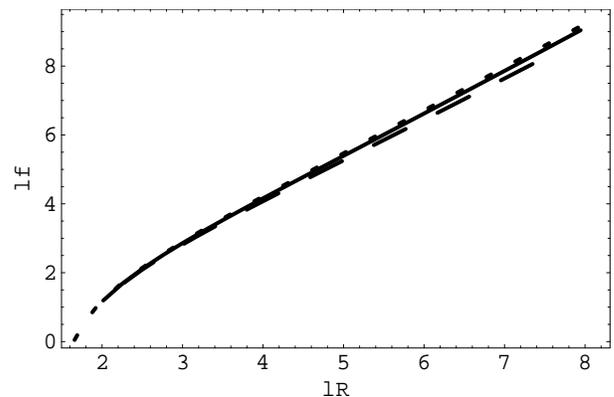}}
\caption{Reconstructed $f(R)$ for Chaplygin gas with $\Omega_M = 0.3$   and three different values of the normalization constant, namely $A_s = 0.25$ (short dashed), $A_s = 0.50$ (solid) and $A_s = 0.75$ (long dashed). We report $lf = \ln{(-f)}$ as function of $lR = \ln{(-R)}$ rather than $f(R)$ to better show the results.}
\label{fig: fvsrcg}
\end{figure}

We may apply our procedure to recover the $f(R)$  theory which reproduces the same dynamics of the GCG. For simplicity, we concentrate our attention only to the case $\alpha = 1$ that is the originally proposed Chaplygin gas. The Ricci scalar turns out to be\,:

\begin{equation}
R = - 3 H_0^2 \left [ {\cal{C}}(z) - \frac{3 (1 - A_s) (1 + z)^6}{{\cal{C}}(z)} \right ]
\label{eq: rvszcg}
\end{equation}
with\,:

\begin{equation}
{\cal{C}}(z) = \sqrt{A_s + (1 - A_s) (1 + z)^6} \ .
\label{eq: defc}
\end{equation}
Proceeding as for the quiessence case, we get the following equation for $f(z)$\,:

\begin{equation}
{\cal{C}}_3(z) \frac{d^3f}{dz^3} + {\cal{C}}_2(z) \frac{d^2f}{dz^2} + {\cal{C}}_1(z)\frac{df}{dz} = -3 H_0^2 \Omega_M (1 + z)^3
\label{eq: eqcg}
\end{equation}
where ${\cal{C}}_i(z)$ are  functions of the redshift $z$ and $A_s$  which are derived by inserting Eq.(\ref{eq: hcg}) into Eqs.(\ref{eq: h1})\,-\,(\ref{eq: ddotR}). The numerical solution of Eq.(\ref{eq: eqcg}) is plotted in Fig.\,\ref{fig: fvszcg}, while Fig.\,\ref{fig: fvsrcg} shows the corresponding $f(R)$ obtained by eliminating $z$ using Eq.(\ref{eq: rvszcg}). An important caveat is in order here. Although $\Omega_M$ is identically $1$ in the usual approach to Chaplygin gas, one has to choose a value of $\Omega_M \neq 1$ since the Friedmann equations have been modified. Moreover, this choice should be based on model\,-\,independent determination of this parameter which also enters the determination of the boundary condition $f(z = 0)$ through Eq.(\ref{eq: fzero}). We set $\Omega_M = 0.3$ as for the best fit concordance $\Lambda$CDM model since this value is also in agreement with the model independent estimates coming from the abundance of galaxy clusters.

The similarity among the curves for different values of $A_s$ is striking, but not fully unexpected. Actually, at larger $z$ (where the curves are almost overlapping in these logarithmic plots), the Chaplygin gas reduces to a matter only universe whatever is the value of $A_s$ so that the dependence on this parameter is washed out. A stronger dependence on $\alpha$ may be expected in the GCG scenario since this parameter controls the rate of transition from a $\Lambda$ dominated to a matter dominated universe.

Finally, we have checked that the numerical solution for $f(R)$ is very well approximated by Eq.(\ref{eq: fit}) with\,:

\begin{figure}
\centering \resizebox{8.5cm}{!}{\includegraphics{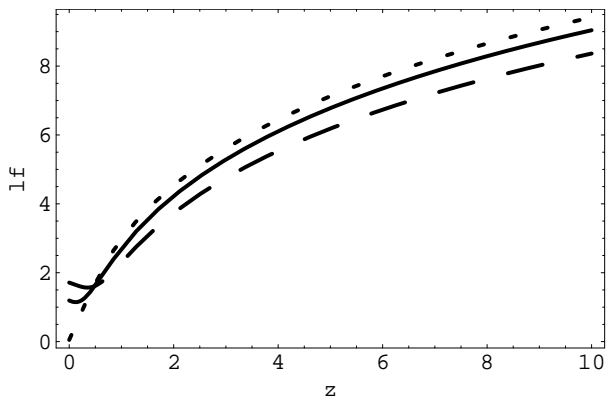}}
\caption{Reconstructed $f(z)$ for Chaplygin gas with $\Omega_M = 0.3$   and three different values of the normalization constant, namely $A_s = 0.25$ (short dashed), $A_s = 0.50$ (solid) and $A_s = 0.75$ (long dashed). We report $lf = \ln{(-f)}$ rather than $f$ to better show the results.} 
\label{fig: fvszcg}
\end{figure}

\begin{displaymath}
(l_1, l_2, l_3, l_4) = (1.9814, 0.5558, 0.2665, -2.5337)
\end{displaymath}
for the model with $(\Omega_M, A_s) = (0.3, 0.75)$.  We stress again that Eq.(\ref{eq: fit}) is only an empirical fitting function so that it is dangerous to draw any physical implications from the fact that the same functional expression for $f(R)$ works well for both quiessence and Chaplygin gas models. Actually, such a result could be somewhat explained noting that both classes of models are designed to fit the same dataset over the same redshift range. As such, there is a sort of degeneracy among quiessence and Chaplygin gas that could be the reason why Eq.(\ref{eq: fit}) works well for both class of dark energy models.

\section{Discussion and Conclusions}

The observed cosmic acceleration  may be seen as the first signal of a breakdown of the Einstein theory of General Relativity. Indeed, several theoretical motivations claims for modifications of General Relativity. Motivated by these considerations, much attention has been recently devoted to higher order theories of gravitation which are obtained by replacing the Ricci scalar $R$ with a generic function $f(R)$ in the gravity Lagrangian. Usually referred to as $f(R)$ theories, these scenarios make it possible to explain the observed cosmic acceleration without introducing any scalar field or changing the properties of  matter term (curvature quintessence). Although physically well motivated and mathematically elegant, this approach has its own problems. The metric formulation of $f(R)$ theories leads to a fourth order nonlinear differential equation for the scale factor $a(t)$ which cannot be analytically solved in general even for some simple choices of the function $f(R)$. Moreover, a numerical solution, although possible, is difficult to handle because of the large uncertainties on the parameters $(a_0, H_0, q_0, j_0)$ which determine the boundary conditions.

It is worth noting that the  same equation may also be seen  as a linear third order differential equation for the function $f(z)$. For a given $H(z)$, this equation is easier to solve numerically since the boundary conditions may be set on the basis of physical considerations only. We have therefore developed a straightforward procedure which makes it possible to solve for $f(R)$, given an expression of the Hubble parameter $H(z)$ and a physically motivated choice of the boundary conditions. We stress that our approach to $f(R)$ theories is reversed with respect to the usual one. Rather than arbitrary choosing the function $f(R)$ and then compare the corresponding model to the observational data, we reconstruct $f(R)$ from a quantity which (at least, in principle) may be directly estimated from the data. As a consequent result, the $f(R)$ theory so obtained will intrinsically fit the data and thus it is a viable candidate to solve the dark energy puzzle.

Our method to  determine $f(R)$ needs a precise knowledge of the Hubble parameter $H(z)$. This quantity could be derived, for instance, from the luminosity distance of SNeIa, but the uncertainties are still too large thus preventing a full reconstruction of $f(R)$ directly from the data. An easy way to overcome this problem is resorting to a parametrized expression of $H(z)$ to be fitted against the available data in order to determine its parameters. This function may then be used as input for the pipeline we have devised. We have explored one possible choice also evaluating (the order of magnitude of) the uncertainty on $R(z)$. Exploring in detail this approach, however, demands for simulated datasets in order to estimate the error on the reconstructed $f(R)$ and the properties of the sample needed for a reliable determination of the gravity Lagrangian directly from the data. This further issue will be investigated in a forthcoming paper.

Although implemented  as an {\it observations\,-\,based} method, it is possible to use our procedure as a simple way to investigate {\it intrinsic} degeneracies among popular dark energy models and $f(R)$ theories. We have indeed shown that from the point of view of dynamics of the late time universe, higher order theories of gravity may mimic several dark energy models. Actually, using as input for our pipeline the Hubble parameter corresponding to a given model, we may found the $f(R)$ theory which reproduces the same dynamics and thus fit the data with the same goodness as the given dark energy model. We have explicitly shown this for quiessence (dark energy with constant equation of state) and Chaplygin gas (as a prototype of UDE models). It is worth noting that the same functional expression fit well the numerically reconstructed $f(R)$ for both class of models. Although unexpected, this result could also be a consequence of the fact that both models fit the same dataset so that they are forced to resemble each other over the redshift range probed by the data. However,  the corresponding $f(R)$ theories extend this similarity over the full evolutionary history of the universe since it is likely that the functional shape of $f(R)$ does not change with $z$ even if $R$ evolves. The successful application of the method to two radically different dark energy models is perhaps the most important result of this paper since it suggests that the crowded zoo of dark energy models may be seen as the result of our ignorance of what is the correct expression for $f(R)$ that has to be inserted in the gravity Lagrangian. Actually, observations tell us that something is wrong with the old standard picture of the universe, but cannot indicate what it is. It is then somewhat a matter of taste to choose whether we are lacking an ingredient (such as a scalar field), or the matter equation of state need to be modified (as in UDE models) or rather it is the underlying theory of gravity which has to be generalized. What we have shown here is that these three philosophycally different approaches may indeed be considered as distinct manifestations of the same physics of the late universe. 

Discovering what is this  physics remains an open problem demanding for both theoretical investigations and observational evidences. From an observer's point of view, CMBR anisotropy and polarization spectra and the clustering properties of the large scale distribution of galaxies are ideal tools to discriminate among the rival possibilities outlined above. Indeed, while the SNeIa Hubble diagram and the data on the gas mass fraction in galaxy clusters probe only the Hubble parameter, both CMBR and large scale structure depend on how the evolution of perturbations takes place in the background cosmological model. Actually, the perturbation equations for scalar field quintessence and UDE models are radically different from those in $f(R)$ theories even if the dynamics of the model (controlled by the Hubble parameter) is the same. From a theoretician's point of view, choosing among higher order gravity theories and dark energy models is quite difficult because of the intrinsic degeneracy we have shown. Nonetheless, constraints on $f(R)$ theories may be obtained by considering the full evolutionary history of the universe. Since the functional expression of the gravity Lagrangian does not change with $z$, a given $f(R)$ theory should not only describe the late time universe, but also give rise to an inflationary period. If we will be able to reconstruct $f(R)$ directly from the data without resorting to a parametrized expression, we could study the picture of the universe assigned by this model and discard such $f(R)$ theories which are  not able to reproduce inflation.

The procedure we have presented relies  on the field equations (\ref{eq: field}) that have been obtained varying the Lagrangian with respect to the metric only. Actually, as we have discussed above,  $f(R)$ theories may also be studied using the Palatini approach where the variation is performed with respect to the metric and the connection considered as independent variables. Being the field equations different, the procedure we have implemented does not hold in the Palatini formulation of $f(R)$ theories. We note, however, that it is still not clear what is the correct approach to higher order gravity theories. Indeed, the possibility of mimicking dark energy models offered by the metric formulation of $f(R)$ theories could be considered as an encouraging evidence favouring this strategy because of the capability to reproduce all the successful dark energy phenomenology. A similar method has to be developed also for the Palatini formulation, as we are going to do in \cite{ACCFT}. 

We would like to conclude with a general comment. The method we have presented makes it possible to reconcile dark energy models and $f(R)$ theories as two different faces of the same medal. Both rival theories are able to reproduce the same dataset and thus to describe the same dynamics of the late time universe since we have shown that they are distinct manifestations of the same underlying physics. Discriminating among them is only possible going back into the past to the period of structure formation and still before, up to inflation. In our opinion, a unified study of both the early and late universe is the unique way to understand what is the correct physics. Higher order gravity theories naturally offer this possibility and may thus be used as powerful light to look into the dark sector of the universe.

\appendix*

\section{Stability of $f(R)$ theories}

Soon after the first proposals of $f(R)$ theories as alternative explanations of the cosmic acceleration, several criticisms were raised on the stability of such models in the metric formulation. In particular, in Ref.\,\cite{DK03}, it was pointed out that the model $f(R) = R + \mu^4/R$ suffers violent instabilities and was argued that something similar should hold in every theory of gravity which leads to higher order differential equations for the scale factor. This argument has been often used as an evidence against the metric formulation of $f(R)$ theories thus motivating the great interest dedicated to the Palatini approach which avoids this problem. Actually, things are different. It is, indeed, possible to show that a leading role in the determination of the stability of the theory is played by the following potential (see, e.g., \cite{Nojiri})\,:

\begin{eqnarray}
U(R_0) & \equiv & \left [ \frac{f^{(4)}(R_0)}{f^{(2)}(R_0)} - \frac{f^{(3)}(R_0)^2}{f^{(2)}(R_0)^2} \right ]
\nabla_{\alpha} R_0 \nabla^{\alpha} R_0 + \nonumber \\
~ & ~ & + \frac{R_0}{3} - \frac{2 f^{(1)}(R_0) f^{(3)}(R_0) R_0}{3 f^{(2)}(R_0)^2} \nonumber \\
~ & ~ & - \frac{f^{(1)}(R_0)}{3 f^{(2)}(R_0)} + \frac{2 f(R_0) f^{(3)}(R_0)}{3 f^{(2)}(R_0)^2} \nonumber \\
~ & ~ & - \frac{R_0 f^{(3)}(R_0)}{3 f^{(2)}(R_0)^2}
\label{eq: stabpot}
\end{eqnarray}
where $f^{(i)} = d^if/dR^i$ and $R_0$ is the solution of the unperturbed field equations. Note that, to be coherent with Ref.\,\cite{Nojiri}, here we are using the signature $\{-, +, +, +\}$ so that $R_0$ is positive and the trace of the matter stress\,-\,energy tensor is negative. Eq.(\ref{eq: stabpot}) has been obtained by setting $R = R_0 + R_1$ with $|R_1| << |R_0|$ and developing the equation at the first perturbative order. If $U(R_0)$ is positive, since $\Box R_1 \sim -\partial_t^2 R_1$, the perturbation $R_1$ becomes exponentially large and the system is unstable. Assuming that the matter is uniformly distributed, we may simplify Eq.(\ref{eq: stabpot}) setting $\nabla_{\alpha} R_0 = 0$ and then study the sign of $U(R_0)$ for a given $f(R)$ theory.

For the model $f(R) = R + \mu^4/R$, $U(R_0)$ turns out to  be positive so that the theory is indeed unstable. But this is not a general result. For instance, the choice $f(R) = \beta R^n$ \cite{capozcurv,review,curvfit} gives\,:

\begin{equation}
U(R_0) = \frac{(n - 2) (2 \beta R_0^{n - 1} - 1)}{3 \beta n (n - 1) R_0^n} \ .
\label{eq: stabrn}
\end{equation}
Assuming that the coupling constant $\beta$ is positive, the constraint $U(R_0) \le 0$ reduces to\,:

\begin{eqnarray}
2 \beta R_0^{n - 1} - 1 \ge 0 & {\rm for} & \ n \le 0 \ {\rm and} \ 1 \le n \le 2 \ , \nonumber \\
~ & ~ & ~ \nonumber \\
2 \beta R_0^{n - 1} - 1 \le 0 & {\rm for} & \ 0 \le n \le 1 \ {\rm and} \ n > 2 \ , \nonumber
\end{eqnarray}
so that the stability of the theory depends on $n$ and $\beta$. A similar analysis can be conducted for the approximating function (\ref{eq: fit}) in order to determine the stability of the reconstructed $f(R)$. It is thus worth stressing, as final remark, that the usual criticism against the metric approach to $f(R)$ theories about the stability arguments has to be reconsidered on a case by case basis. Therefore, we are confident that the procedure we have implemented to reconstruct $f(R)$ is meaningful and not affected by any systematic problem related to the metric formulation of $f(R)$ gravity.

\end{document}